\documentclass[preprints,article,accept,moreauthors,pdftex]{Definitions/mdpi}
\firstpage{1} 
\pubvolume{1}
\issuenum{1}
\articlenumber{0}
\pubyear{2022}
\copyrightyear{2022}
\datereceived{} 
\dateaccepted{} 
\datepublished{} 
\hreflink{https://doi.org/} 

\Title{Numerical simulation using cellular Potts model for wound closure with ATP release and the mechanobioligical effects}

\TitleCitation{Title}

\Author{Kenta Odagiri $^{1,2}$\orcidA{}, Hiroshi Fujisaki $^{2,3}$\orcidB{},
Hiroya Takada $^{2,4,5}$ and Rei Ogawa $^{2,5}$}

\AuthorNames{Kenta Odagiri, Hiroshi Fujisaki, Hiroya Takada and Rei Ogawa}

\AuthorCitation{Odagiri, K.; Fujisaki, F.; Takada, H.; Ogawa, R.}

\address{%
$^{1}$ \quad School of Network and Information, Senshu University,
2-1-1, Higashi-Mita, Tama, Kawasaki, Kanagawa~214-8580, Japan; k-oda@isc.senshu-u.ac.jp\\
$^{2}$ \quad AMED-CREST, Japan Agency for Medical Research and Development, 1-1-5 Sendagi, Bunkyo-ku, Tokyo~113-8603,~Japan\\
$^{3}$ \quad Department of Physics, Nippon Medical School, 1-7-1 Kyonan-cho, Musashino, Tokyo 180-0023, Japan; fujisaki@nms.ac.jp\\
$^{4}$ \quad Department of Anti-Aging and Preventive Medicine, Nippon Medical School, 1-1-5 Sendagi, Bunkyo-ku, Tokyo~113-8603,~Japan; h-takada@nms.ac.jp\\
$^{5}$ \quad Department of Plastic, Reconstructive and Aesthetic Surgery, Nippon Medical School, 1-1-5 Sendagi, Bunkyo-ku, Tokyo~113-8603,~Japan; r.ogawa@nms.ac.jp\\
}

\corres{Correspondence: k-oda@isc.senshu-u.ac.jp; Tel: +81-44-900-7915}

\abstract{
To computationally investigate the recent experimental finding 
such that extracellular ATP release caused by exogeneous mechanical 
forces promote wound closure, 
we introduce a mathematical model, the Cellular Pots Model (CPM), 
which is a popular discretized model on a lattice, 
where the movement 
of a "cell" is determined by a Monte Carlo procedure. 
In the experiment, it was observed that 
there is mechanosensitive ATP release from 
the leading cells facing the wound gap
and the subsequent intracellular Ca$^{2+}$ influx.
To model these phenomena, the Reaction-Diffusion equations 
for ATP and Ca$^{2+}$ concentrations
are adopted and combined with CPM, where we 
also add a polarity term because the cell migration is 
enhanced in the case of ATP release.
From the numerical simulations using this hybrid model,
we discuss effects of the collective cell migration due to 
the ATP release and the Ca${}^{2+}$ influx
caused by the mechanical forces and 
the consequent promotion of wound closure.
}


\begin{document}

\section{Introduction}

Recent studies in the field of mechanobiology 
\cite{mechanobiobook1,mechanobiobook2,mechanobiobook3} have
revealed that
cells recognize and respond to various external stimuli
or external forces,
and it has been found that 
various biological functions in vivo are 
triggered by this mechanobioligical effects
\cite{Hayakawa, Ingber, Jansen1,Ladoux,Huang, He}.
As such, mechanobiological issues are
important and challenging not only for basic biology
but also for the optimal treatment of plastic surgery 
which is called "mechanotherapy"
\cite{Thompson, Akaishi, Ogawa, Ogawa2, Ogawa3}.

In relation to wound closure and mechanobiology,
Takada, Furuya and Sokabe 
found that the mechanical forces on cells are 
important for wound closure \cite{Takada1}.
In their experiment, 
a linear wound was made in cultured vascular endothelial cells,
and the wound closure process was examined by applying stretch stimulation.
It was found that the wound closure rate increased 
with the stretch stimulation compared to no external stimulation.
They also measured real-time imaging of ATP and Ca${}^{2+}$ 
concentrations,
where the stretch stimulation releases 
large amounts of ATP from the cells closest to the wound gap.
Here ATP functions as an extracellular messenger and increases intracellular Ca${}^{2+}$ levels.
When ATP travels like a transient wave from the proximal wound to the posterior cell,
it is followed by a transient wave of Ca${}^{2+}$.
It was shown that the response of the cells to Ca${}^{2+}$ was indispensable for the wound closure
because the Ca${}^{2+}$ wave disappeared and cell migration did not occur
when the extracellular Ca${}^{2+}$ was removed.

Motivated by this experimental study, 
we here propose a mathematical model of
wound closure process that explicitly considers ATP release by mechanical stimulation to the cells,
the accompanying change in Ca${}^{2+}$ concentration,
and collective cell migration by response to Ca${}^{2+}$.
We construct a hybrid model of the Cellular Potts Model (CPM)
\cite{Glazier1, Glazier2, Merks, Hirashima, Guisoni, Czirok, Rens1} 
and the Reaction-Diffusion (RD) equations \cite{Kapral, Kuramoto},
where CPM represents the dynamics of cells, such as cell migration and cell growth,
and RD represents the changes in ATP and Ca${}^{2+}$ concentration.
The cell migration in response to Ca${}^{2+}$ is expressed as self-propulsion driven by cell polarity, which is further 
added as an energy term in the CPM.
Using this hybrid model, we try to numerically 
simulate the experiment on wound closure by 
Takada and coworkers.

This paper is organized as follows.
In Sec. II, we first introduce the hybrid 
mathematical model for the wound closure process.
We next show the numerical results of our model in Sec. III.
To compare with the experimental results such that 
the wound closure is promoted by applying the mechanical stimulation,
we here examine two different scenarios with and without ATP release.
We also discuss why the wound closure rates are different 
in these two cases. Finally, we conclude this paper in Sec. IV.

\section{Computational Model}

We here introduce a mathematical model of cellular dynamics for wound closure, taking into account the intracellular ATP release and the subsequent extracellular Ca${}^{2+}$ influx
by the mechanical stimulation.
To couple the cellular dynamics 
with the spatio-temporal dynamics of the chemical components
inside and outside of the cells,
we propose a hybrid model of the Cellular Potts Model and the Reaction-Diffusion Equations.
The former model represents the cellular population dynamics and the latter represents the spatio-temporal
dynamics of ATP and Ca${}^{2+}$, and the details 
are explained below.

\subsection{Cellular Dynamics}

Collective cell migration is one of the key processes 
in the wound closure process,
and various mathematical models have been proposed
\cite{Palachanis, Suzuki, Barton, Ishihara, Vicsek}.
We here employ the Cellular Potts Model (CPM) for the computational modeling of the cellular collective movement.
CPM is defined on a lattice, representing a cell shape 
as a group of sites on the lattice.
Each site has a cell index $\sigma$, 
where the same cell shares the same cell index $\sigma$.
To mimic various cellular behavior such as cell 
deformation and migration,
CPM repeats the "exchange" of adjacent lattice sites 
with a Metropolis algorithm using 
a kind of "energy" as described below.

The move is described as an attempt to copy the cell index $\sigma ( \vec{x} )$
at a randomly selected site $\vec{x}$ into another
randomly selected neighboring site $\vec{x'}$.
Whether the move is accepted or not 
is determined using
the change in the total energy $\Delta E$ due to the exchange.
When $\Delta E \leq 0$, 
the move is always accepted, 
however, when $\Delta E > 0$,
the move is accepted with the 
following probability:
$p( \vec{x} \to \vec{x'} )$ $= \exp (- \beta \Delta E)$
where $\beta$ denotes the magnitude of cell fluctuation, 
in accord with the conventional Metropolis algorithm.

\subsection{Energy of Cell Population}

The energy of cell population used in the 
CPM calculations is modelled 
as follows:
\begin{eqnarray}
\label{eq:CPM}
 E_{\rm CPM} &=& 
\sum_{\vec{x},\vec{x'}}J_{\tau{\left( \sigma \left( \vec{x} \right) \right)},
\tau{\left( \sigma \left( \vec{x'} \right) \right)}}
\left( 1 - \delta_{\sigma \left( \vec{x} \right), \sigma \left( \vec{x'} \right)}  \right) \notag \\
&+& \sum_k \lambda_V ( V_k - V_{t,k})^2 + \sum_k \lambda_S ( S_k - S_{t,k})^2
- E_{\text{mig}},
\end{eqnarray}
where the first term represents the interfacial energy due to the cell adhesion, $\vec{x}$ and  $\vec{x'}$ denote two neighboring lattice sites,
$\sigma ( \vec{x} )$ denotes a cell index at a lattice site $\vec{x}$, and
$\tau$ denotes a cell type. (Note that the cell index refers to a single cell, whereas the cell type does bunches of cells.)
$J_{\tau,\tau'}$ denotes the adhesion energy between cell type $\tau$ and $\tau'$.
The second term represents cell volume constraint energy, where
$V_k$ and $V_{t,k}$ denote a calculated volume and a target volume of cell $k$, respectively,
and $\lambda_V$ is the strength of the volume constraint.
The third term represents cell surface constraint energy, where
$S_k$ and $S_{t,k}$ denote a calculated surface and a target surface of cell $k$, 
respectively, and $\lambda_S$ is the strength of the surface constraint.
(In two-dimensional system, cell surface area means the length of cell's perimeter.) 
The last term  represents a cell migration energy
due to cellular self-propulsion as described below.

\subsection{Cellular Self-Propulsion}
The cell migration energy $E_{\text{mig}}$
with the self-propulsion is given by
\begin{eqnarray}
\label{eq:EmigCP}
E_{\text{mig}} = \sum_k P \frac{\vec{p_k}}{\left| \vec{p_k} \right|}
\cdot \vec{v_k},
\end{eqnarray}
where
$\vec{p_k}$ represents the cell polarity vector of cell $k$
and $\vec{v_k}$ represents the displacement vector of the center 
of  cell $k$
due to the Monte Carlo move.
$P$ is the strength of the cell autonomous motility.
This formulation indicates that the cell polarity, which reflects the spatial differences in
the biochemical state of a cell, drives cellular self-propulsion.
Szab\'{o} and Czir\'{o}k proposed that the cell polarity vector $\vec{p_k}$ decays spontaneously
but reinforced by cell displacements \cite{Szabo1,Szabo2}.
We here assume that Ca${}^{2+}$ ions enhance the cellular self-propulsion and
therefore the cell polarity vector is also reinforced in proportion to the amount of Ca${}^{2+}$.
In each Monte Carlo step (MCS), the change in $\vec{p_k}$ is thus given by
\begin{eqnarray}
\label{eq:dpk}
\Delta \vec{p_k} = -\gamma \vec{p_k} + \Delta \vec{g_k}
+ \lambda_{c} \sum_{\vec{x_{k}}}  W \left( C \left( \vec{x_{k}} \right) \right) \left( \vec{x_{k}} - \vec{g_{k}} \right).
\end{eqnarray}
The first term indicates spontaneous decay of $\vec{p_k}$ with decay rate $\gamma$.
The second term indicates reinforcement of $\vec{p_k}$ resulting from cell displacements, where
$\vec{g_k}$ and $\Delta \vec{g_k}$ are the center of cell $k$ and the change in $\vec{g_k}$ during the MCS.
The last term indicates reinforcement of the cell polarity due to intracellular Ca${}^{2+}$ ions,
where 
$C(\vec{x})$ is the concentration of Ca${}^{2+}$ ion at position $\vec{x}$ in cell $k$, 
$W( C(\vec{x_{k}}) ) = \exp \left( -\frac{1}{w C(\vec{x})} \right) $ is the weight function of
$C(\vec{x_{k}})$, $w$ is the weight coefficient of $C(\vec{x_{k}})$, and
$\lambda_{c}$ is the magnitude of Ca${}^{2+}$ ion's enhancement of $\vec{p_{k}}$.

\subsection{Dynamics of Chemical Components}
The change in the concentration of extracellular ATP $A(\vec{x},t)$
and intracellular Ca${}^{2+}$ $C(\vec{x},t)$ are modelled by the following Reaction-Diffusion equations.
\begin{eqnarray}
\label{eq:dtA}
\frac{\partial A(\vec{x},t)}{\partial t} &=&
\delta_{\sigma(\vec{x}),\sigma_{\text{Edge}}} \theta (t) \theta (t_b-t) k_0
- \delta_{\sigma(\vec{x}),\sigma_{\text{cells}}} k_{A} A(\vec{x},t) + D_C \nabla^2 A(\vec{x},t),\\
\label{eq:dtC}
\frac{d C(\vec{x},t)}{d t} &=& \delta_{\sigma(\vec{x}),\sigma_{\text{cells}}}
\theta (A_{\text{th}}) k_{C} ( C_0 - C(\vec{x},t) ).
\end{eqnarray}
The first and second terms on the right hand side of Equation (\ref{eq:dtA}) 
denote ATP release and ATP consumption.
ATP molecules release from only the leading cells $\sigma_{\text{Edge}}$ that faces the wound edge
at the beginning of the cell migration ($0 \leq t \leq t_b$)
and are consumed by cells $\sigma_{\text{cells}}$ with 
rate $k_A$ as indicated by the second term.
$\theta(t)$ is the Heaviside function (1 when $t>0$ and 0 otherwise), $t_b$ is the time when the ATP release ends,
and $k_0$ is the release rate of ATP.
The last term in Eq.~(\ref{eq:dtA}) denotes 
the extracellular diffusion of ATP with 
the diffusion coefficient $D_C$.
Equation (\ref{eq:dtC}) represents the intracellular Ca${}^{2+}$ response
in the cells $\sigma_{\text{cells}}$ induced by released extracellular ATP molecules, where
$A_{\text{th}}$ is the threshold value of the ATP concentration for
the extracellular Ca${}^{2+}$ influx due to the Ca${}^{2+}$ response,
$k_C$ is the influx rate of Ca${}^{2+}$,
and $C_0$ is the steady-state 
concentration of extracellular Ca${}^{2+}$.

\subsection{Cell Growth and Contact Inhibition}
We further introduce the cell growth 
considering the contact inhibition.
Contact inhibition is the mechanism which normal cells stop proliferation or growth when they contact each other.

Cell growth is modelled by increasing the target volume $V_{t,k}$ and
the target surface area $S_{t,k}$ \cite{Stott,Szabo3} in the CPM.
When the cell $k$ is not completely surrounded by other cells,
the values of $V_{t,k}$ and $S_{t,k}$ increase in a cycle $P_C$,
where the maximum sizes are  $V_{\text{max}}$ and $S_{\text{max}}$, respectively.
On the other hand, when the cell is completely 
surrounded by the other cells,
the values of $V_{t,k}$ and $S_{t,k}$ does not change,
that is, the cell does not get large due to the effect of the contact inhibition.

Incidentally, we do not consider the cell division in this model,
because the time scale of the experimental results \cite{Takada1} is shorter than that of the cell division (a cell does not divide 
within the simulation time).

\section{Results and Discussion}
We here show some numerical results using the 
hybrid model (CPM with polarity + RD equations for chemicals + contact inhibition for cells) 
for the wound closure process.
To mimic the experimental situation \cite{Takada1}, 
we set up the initial distribution of cells
as shown in Fig \ref{fig:1} (a).
The gap region (grey) between the upper and 
lower cells represents the linear wound.
Green cells in Fig \ref{fig:1} (a) indicates the leading cells that face the wound edge.
Only these leading cells release ATP molecules, 
as in the experiment \cite{Takada1}, at the beginning of the simulation (cell migration) until $t_b=1$.
The initial volume of each cell is 25 (5 $\times$ 5) units, 
where we consider a square lattice with 1 $\times$ 1 unit, 
and the initial number of the cells is 288. 
(The total system size is 120 $\times$ 120 units and 
the wound region is the half of it.)
Table \ref{table1} shows the parameters of the numerical simulation (units are arbitrary).
$J_{CC}$ and $J_{CM}$ denote the cell-cell adhesion energy and
the cell-medium (cell free area) adhesion energy, respectively.

To clarify the effect of the collective cell migration due to the ATP release and the subsequent Ca${}^{2+}$ influx resulting from the mechanical forces,
we here numerically examine two different scenarios 
with and without the ATP release: 
The former is referred to as 
ATP model and the latter as ATP-free model.
Compared to our previous study of the wound closure \cite{Odagiri},
we add to this model the effect of contact inhibition
and investigate in detail the effects of the ATP release.
This makes the this study more suitable for comparison
with the previous experimental study \cite{Takada1}.

\begin{table}[htbp]
\centering
\caption{
{\bf Parameters used in the simulations (arbitrary units).}}
\begin{tabular}{cccccccccccc}
\hline
$J_{CC}$ & 9.0 & $J_{CM}$ & 3.0 & $\lambda_V$ & 2.0 & $\lambda_S$ & 0.5
& $\lambda_{c}$ & 5.0 & $P$ & 2.0\\
$\gamma$ & 0.025 & $w$ & 10.0 & $k_0$ & 10.0 & $k_A$ & 0.01 & $D_C$ & 0.1 & $A_{\text{th}}$ & 0.1\\
$k_C$ & 1.0 & $C_0$ & 1.0 & $P_C$ & 10 & $V_{\text{max}}$ & 60 & $S_{\text{max}}$ & 38.7\\
\hline
\end{tabular}
\label{table1}
\end{table}

\subsection{Wound closure by the collective cell migration}

\begin{figure}[H]
\center
\includegraphics[width=13cm]{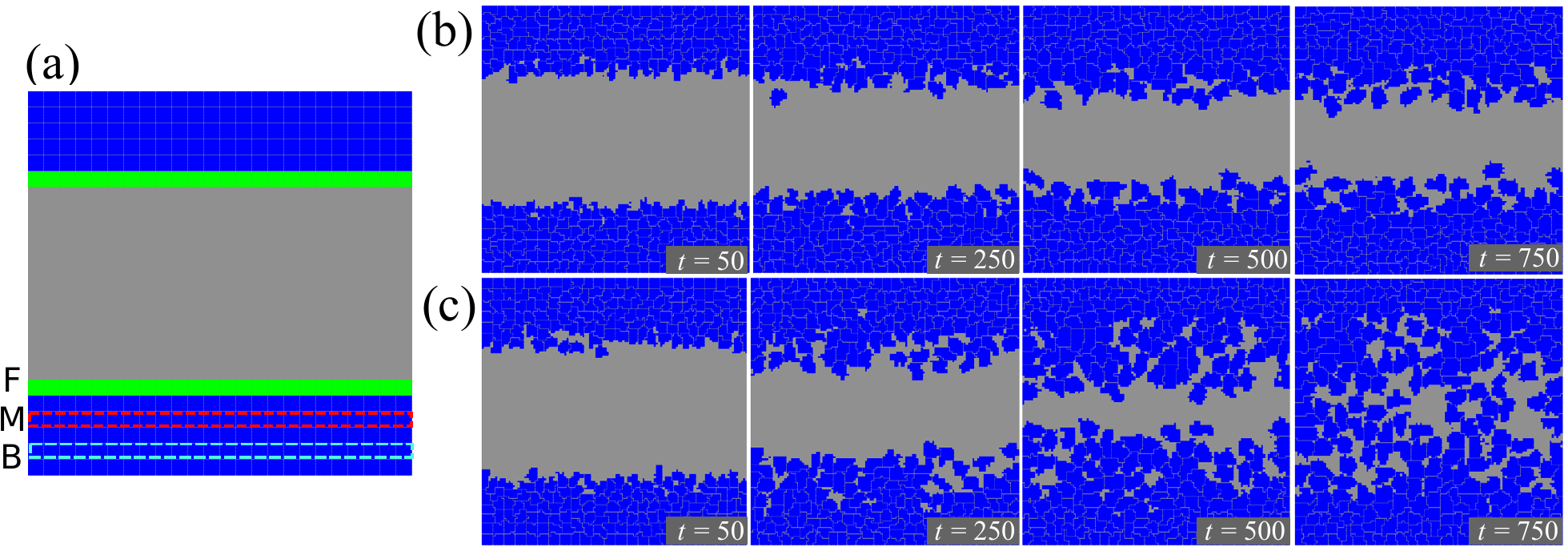}
\caption{(a) Initial distribution of cells.
F, M and B represent cells facing the wound gap, cells located in the middle of the cell cluster
and cells located at the back of the cell cluster, respectively.
(b) and (c) Time evolution of wound closure process in the ATP-free model and the ATP model, respectively.}
\label{fig:1}
\end{figure}

Figure \ref{fig:1} (b) and (c) show the time evolution of the wound closure process
in the ATP-free model and ATP model, respectively.
These figures clearly show that the wound gap closes more quickly in the ATP model than in the ATP-free model.
In the ATP model, the cells facing the wound gap rapidly migrate forward, and this
quick migration causes rapid closure of the wound gap.
On the other hand, in the ATP-free model, the leading cells migrate forward very slowly,
and thus the wound gap closes very slowly.

\begin{figure}[H]
\center
\includegraphics[width=6cm]{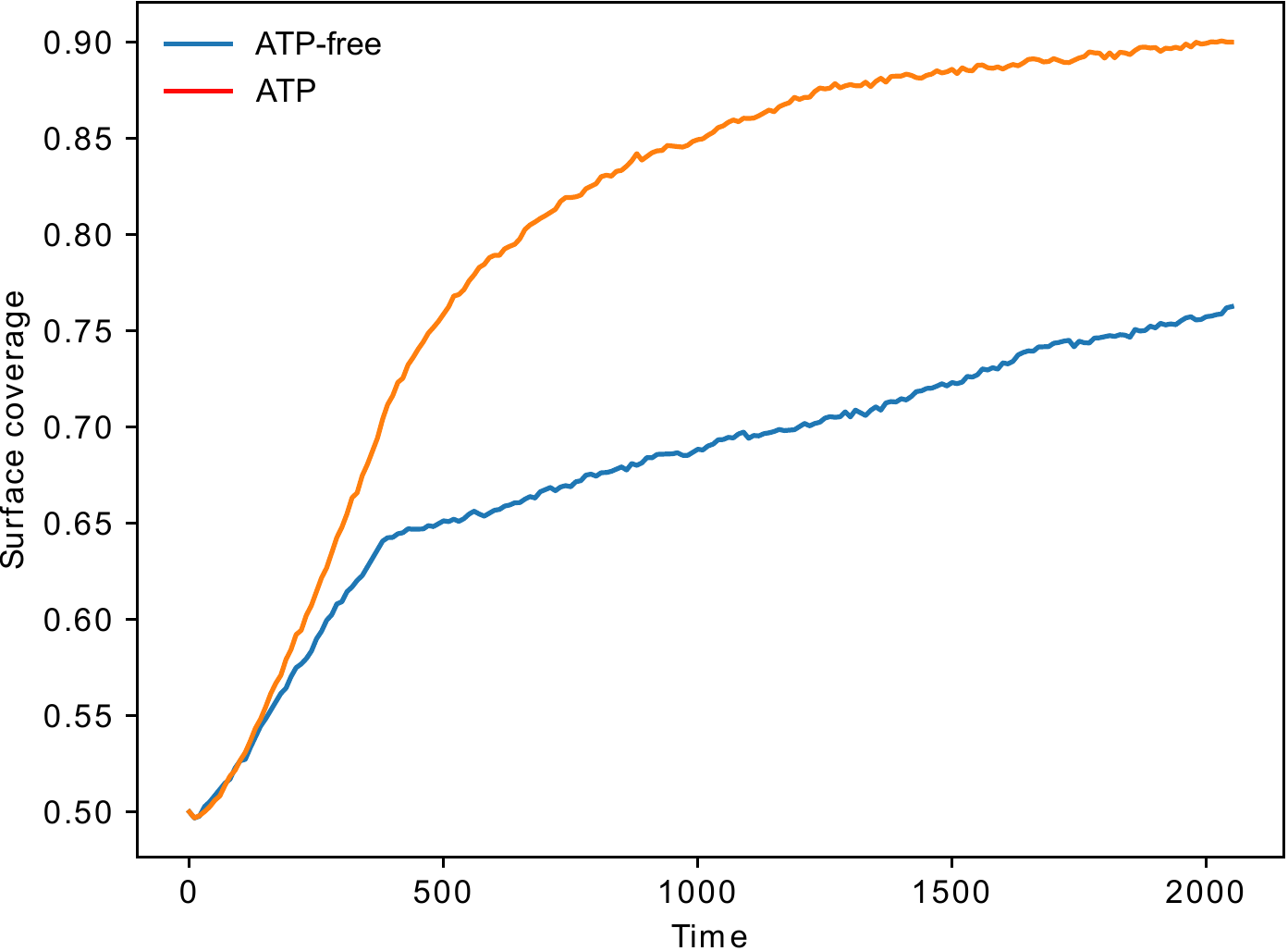}
\caption{Time evolution of the surface coverage by cells in the ATP-free model and the ATP model.}
\label{fig:2}
\end{figure}

To compare the rate of wound closure in these models, we measure the change in
the surface coverage by cells over time (Fig.~\ref{fig:2}).
Although the rate of wound closure in the two cases is almost the same at the early phase ($t=0\sim200$),
the rate in the ATP-free model suddenly slows down after around $t$=400.
On the other hand, in the ATP model,
the wound closure rate does not slow down significantly until cells fill most of the wound gap.
Therefore, the ATP release clearly shows the promotion of the wound closure with the parameters we chose in Table 1.

\subsection{ATP diffusion and Ca$^{2+}$ response}
\begin{figure}[H]
\center
\includegraphics[width=13cm]{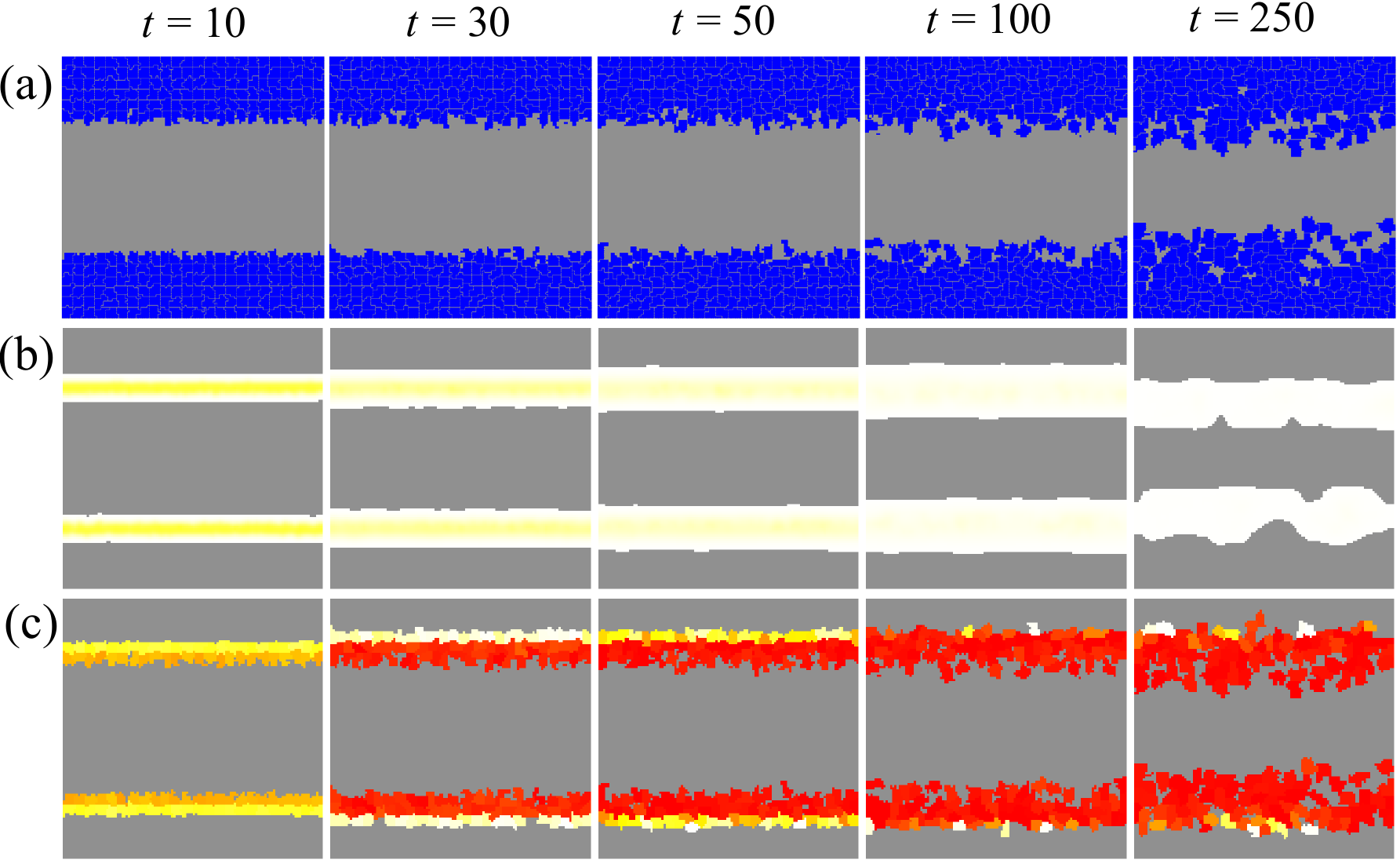}
\caption{Time evolution of the distribution of cells (a), ATP (b)
and Ca$^{2+}$ (c) at the early stage.
Red (white) indicates that the concentration of the chemical components is high (low), and gray indicates that the concentration of them is zero (almost zero).
ATP diffusion and the subsequent Ca$^{2+}$ influx occur quickly although
the wound gap is barely closed.}
\label{fig:3}
\end{figure}

We next show the ATP diffusion and the subsequent Ca$^{2+}$ response at the early stage in the ATP model.
In the previous experimental study \cite{Takada1}, the diffusion of released ATP
and the subsequent Ca$^{2+}$ response have been observed within a few seconds
after mechanical stimulation.
Figure \ref{fig:3} (a), (b) and (c) show the time evolution of the distribution of
cells, ATP and Ca$^{2+}$ at the early stage, respectively.
The released ATP from the leading cells diffuses quickly
and the resulting Ca$^{2+}$ influx occurs
although the width of the wound gap is almost the same ($t=10\sim50$).
And after a while, the diffused ATP decays ($t=100,250$).
These results qualitatively 
correspond to the experimental results.

\subsection{Differences in the wound closure rate}
In order to investigate the difference in the wound closure rate,
we here show the degree of cell migration at various location within the cell cluster
and the distribution of the cell size.

\begin{figure}[H]
\center
\includegraphics[width=13cm]{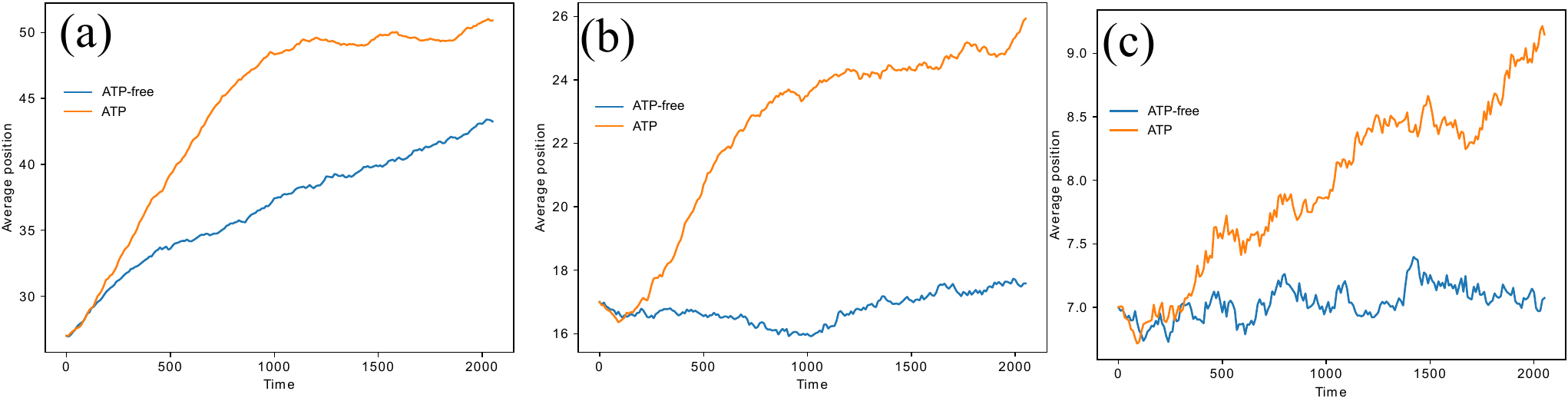}
\caption{(a), (b) and (c) correspond to time evolution of the average cell position for clusters F, M and B, respectively.}
\label{fig:4}
\end{figure}

We first calculate the average cell position (perpendicular to the wound gap) at various location
within the cell cluster, that is,
cells in the row facing the wound gap (cluster F in Fig. \ref{fig:1}(a)),
cells located in the middle row within the cell cluster (cluster M 
in Fig. \ref{fig:1}(a)) and
cells located in the back row within the cell cluster (cluster B in Fig. \ref{fig:1}(a)), respectively.
Figure \ref{fig:4} (a), (b) and (c) show the time evolution of the average cell position for each cluster in two models.
Time evolution of the average position 
for cluster F is quite similar to
that of the surface coverage by 
cells (Fig. \ref{fig:2}).
It indicates that the cell migration of the leading cells directly reflects the wound closure rate.
On the other hand, the average cell positions for clusters 
M and B in the two models show completely different behaviors.
In the ATP model, cells for clusters 
M and B also migrate forward, whereas they remain almost the same positions
in the ATP-free model.
It shows that quick cell migration 
for cluster F due to the ATP release causes
the subsequent cell migration 
for clusters M and B in the ATP model.

\begin{figure}[H]
\center
\includegraphics[width=13cm]{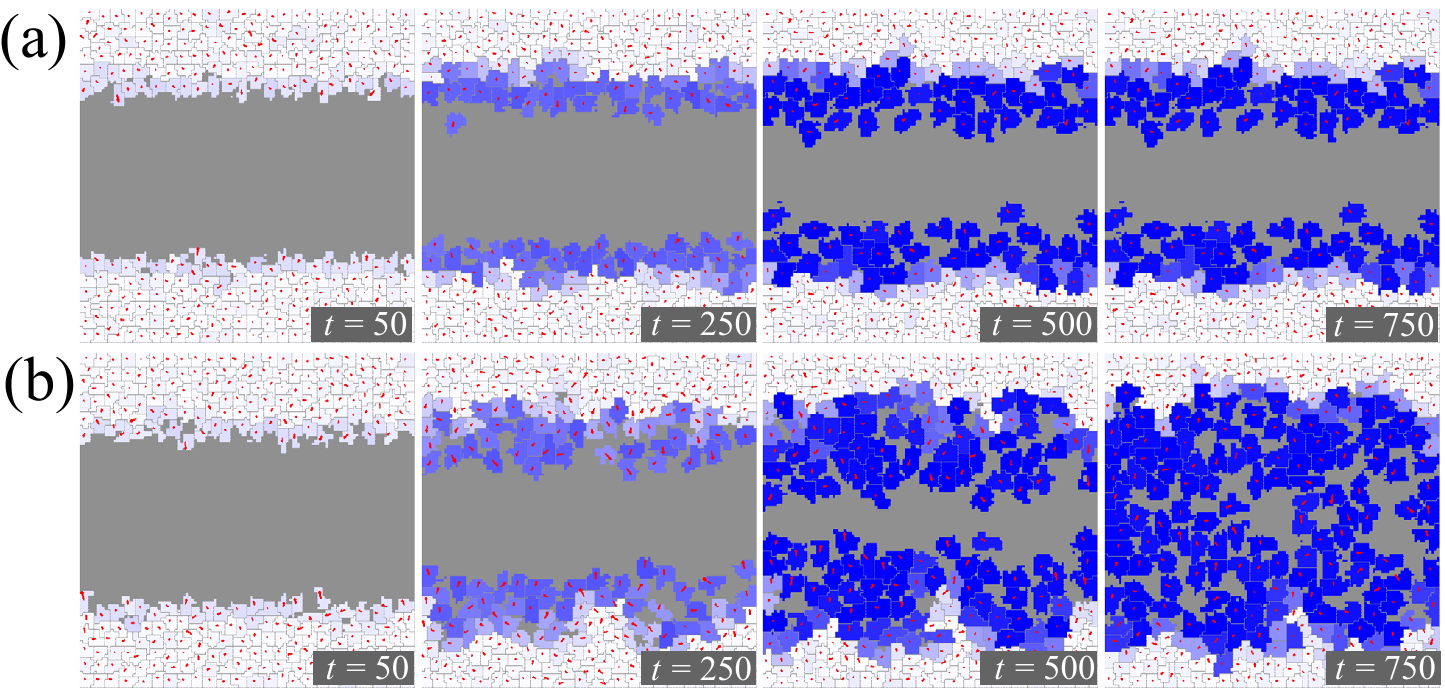}
\caption{(a) and (b) correspond to Fig.~\ref{fig:1}  
(a) and (b), respectively, 
and blue color indicates large cells
whereas white color does small cells.
}
\label{fig:5}
\end{figure}

\begin{figure}[H]
\center
\includegraphics[width=13cm]{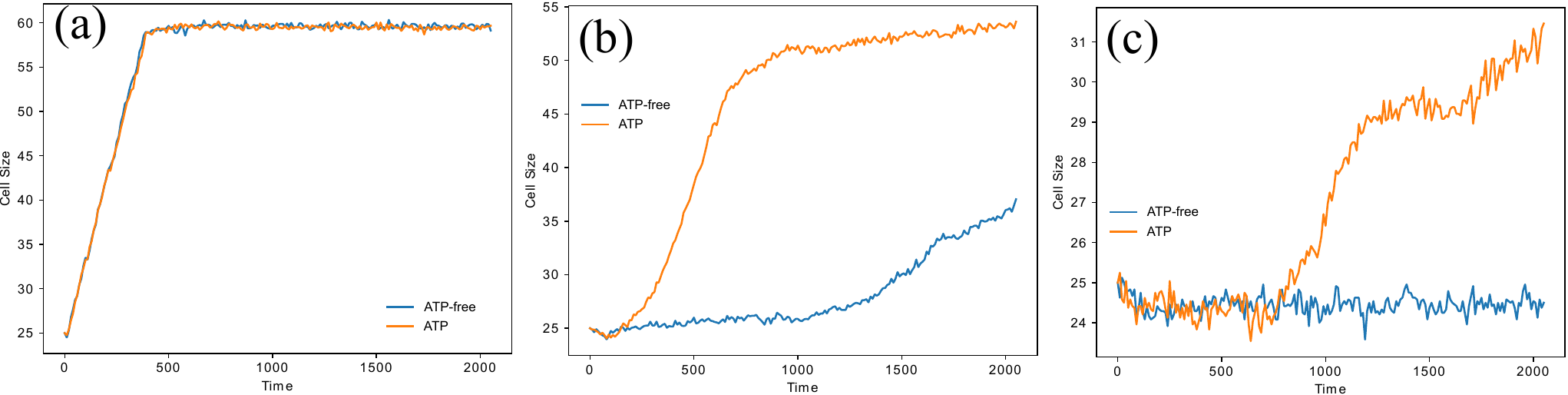}
\caption{(a), (b) and (c) Time evolution of the average cell size for clusters 
F, M and B,
respectively.}
\label{fig:6}
\end{figure}

We next visualize the distribution of the cell size for each cell cluster.
Figure \ref{fig:5} (a) and (b) 
is basically the same as 
Figure \ref{fig:1} (b) and (c) 
but the color code is different: 
Differences in cell color indicate the differences in cell size.
Blue color means the size of a cell is large ($V \simeq V_{\text{max}}$),
and white color means the size of a cell is small ($V \le 25$).
Figure \ref{fig:5} (a) clearly 
indicates that only the cells for cluster F grow and the size of the cells 
for cluster M and B (far from the wound)
hardly changes in ATP-free model.
To investigate it quantitatively, we calculate the average cell size at various locations for clusters F, M and B.
Although cells for cluster F in two models have almost the same size over time (Fig. \ref{fig:6} (a)),
cells at M and B in the ATP model grow faster than those in the ATP-free model  (Fig. \ref{fig:6} (b) and (c)).
This is the reason why the increase in the surface coverage slows down in the ATP-free model.

\begin{figure}[H]
\center
\includegraphics[width=13cm]{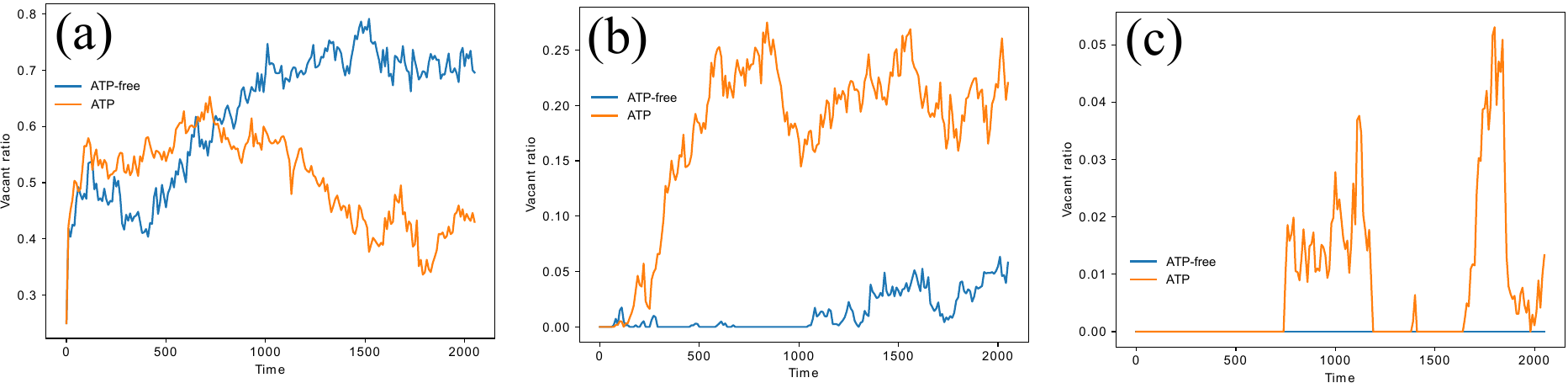}
\caption{(a), (b) and (c) Time evolution of the average vacant ratio for clusters F, M and B, respectively.}
\label{fig:7}
\end{figure}

We next investigate why cells inside the cluster in the ATP-free model can hardly grow.
In our numerical model, a cell surrounded by other cells cannot get large due to the contact inhibition.
Therefore, vacant ratio of a cell, which denote how much space is available around the cell, is a useful
for determining whether the cell can grow or not.
Figure \ref{fig:7} (a), (b) and (c) show the time evolution of the average vacant ratio for clusters F, M and B,
respectively.
In ATP-free model, cells inside the cluster (clusters M and B) are almost completely surrounded 
by the other cells (average vacant ratio is almost zero),
and thus they cannot get large due to the contact inhibition.
Incidentally, the decrease in the vacant ratio for cluster F in the ATP model after $t=700$ is due to
the rapid migration and growing of the leading cells and the resulting loss of vacant space.

From the above considerations, 
we summarize the reason for the promotion of wound closure in the ATP model as follows.
The cells inside the cluster also grow in the ATP model,
and rapid migration of cells on the surface makes space for the cells located away from the wound gap,
resulting in the cell enlargement.
Therefore, the collective cell migration induced by the ATP release and the Ca${}^{2+}$ influx
promotes the cell growth inside the cell cluster and the consequent fast wound closure.

\section{Conclusions}
We have introduced a hybrid model which consists of the CPM (cell dynamics) and the RD equations
(dynamics for chemical components) for the wound closure process, and further added the cell polarity and contact inhibition.
We here focused on the wound closure rate for the ATP-free model (without ATP release from the cells facing the wound) and
the ATP model.
The difference in the wound closure rate is caused by the rapid cell migration 
due to the ATP release and the subsequent Ca${}^{2+}$ influx induced
by the mechanical forces, which we model as the polarity of the cell movement.
The rapid cell migration makes space for the cells inside the cell cluster, and consequently
causes rapid growth for these cells.
On the other hand, in the ATP-free model, the leading cells migrate slowly, and thus
the cells inside the cluster do not have enough space to get large
and cannot grow due to the contact inhibition.

Our hybrid model can also be applied to the vascular network formation,
where anisotropic shape of a cell is important to generate such a fibrous form.
By incorporating the effect of cell shape 
and taking into account the effect of the ATP release induced by external forces, 
our hybrid model will be further extended to deal with 
the vascular network formation under mechanobiological conditions.

\vspace{6pt}

\authorcontributions{Conceptualization, K.O, H.F.; methodology, K.O., H.F., H.T., R.O.; validation, K.O., H.F., H.T., R.O.; investigation, K.O., H.F.; data creation, K.O.; writing---original draft preparation, K.O.; writing---review and editing, K.O., H.F., H.T., R.O. 
All authors have read and agreed to the published version of the manuscript.}

\funding{This research was partially supported by the
Japan Society for the Promotion of Science 
(KAKEN 22K11941 to H.F., K.O.,
19K12207 to K.O., H.F., H.T.)
and AMED-CREST, AMED (JP20gm0810012) to K.O., H.F., H.T., R.O.}


\institutionalreview{Not applicable }

\informedconsent{Not applicable}

\dataavailability{Not applicable }

\acknowledgments{We are grateful to Kazue Kudo, Yoshitaro Tanaka, and Shigeyuki Komura 
for inspiring
and useful discussions.}

\conflictsofinterest{The authors declare no conflict of interest.}

\abbreviations{Abbreviations}{
The following abbreviations are used in this manuscript:\\

\noindent 
\begin{tabular}{@{}ll}
CPM & Cellular Potts model\\
RD & Reaction-Diffusion (equations)\\
MCS & Monte Carlo step\\
ATP & adenosine triphosphate
\end{tabular}}
\appendixtitles{no} 



\begin{adjustwidth}{-\extralength}{0cm}

\reftitle{References}




\end{adjustwidth}
\end{document}